\documentstyle [12pt] {article}

\catcode`@=12
\begin{document}
\thispagestyle{empty}
\begin{flushright} {UCRHEP-T129\\July 1994\\}
\end{flushright}
\vspace{0.5in}
\begin{center}
{\Large \bf Unifiable Supersymmetric Left-Right Model\\
with E$_6$ Particle Content\\}
\vspace{1.5in}
{\bf Ernest Ma\\}
\vspace{0.3in}
{\sl Department of Physics\\}
{\sl University of California\\}
{\sl Riverside, California 92521\\}
\vspace{1.5in}
\end{center}
\begin{abstract}
A new supersymmetric gauge model is proposed with particle content chosen
only from the {\bf 27} and {\bf 27*} representations of $\rm E_6$.  The
gauge symmetry $\rm SU(3) \times SU(2)_L \times SU(2)_R \times U(1)$  is
realized at the TeV energy scale and the gauge couplings converge to a
single value at around $10^{16}$ GeV.  A discrete $\rm Z_4 \times Z_2$
symmetry leads to a generalized definition of lepton number and ensures
the absence of tree-level flavor-changing neutral-current interactions
at the electroweak energy scale.
\end{abstract}

\newpage
\baselineskip 24pt

Several years ago, in connection with superstring theory, a supersymmetric
left-right model was proposed\cite{1} which has the interesting property
that the $\rm SU(2)_R$ charged gauge boson $W_R$ has a nonzero lepton
number, together with new exotic quarks $h$ of charge $-1/3$.  Its many
unconventional implications have been studied in a number of subsequent
publications.[2-16]  This model is potentially of great phenomenological
interest, but only if the $\rm SU(2)_R$ breaking scale $\rm M_R$ is low
enough, say of order a few TeV.  However, that poses a problem for the
unification of gauge couplings.  With the particle content of the original
model\cite{1} and the experimentally determined values of the gauge
couplings at the electroweak energy scale, it is simply not possible
for them to converge to a single value unless $\rm M_R$ is very high.
This is a well-known general result for left-right models.\cite{17}
Hence new particles are necessary if unification of the gauge symmetry
is to be achieved.\cite{18}

To be compatible with its possible superstring antecedent, the particle
content of the proposed supersymmetric model in this paper is assumed
to consist of only components
from the {\bf 27} and {\bf 27*} representations of E$_6$.  [The {\bf 27}
of E$_6$ decomposes into {\bf 16} + {\bf 10} + {\bf 1} of SO(10).]  At
the $\rm SU(3) \times SU(2)_L \times SU(2)_R \times U(1)$ level, it is
proposed that there are three copies of
\begin{eqnarray}
Q = (u,d)_L &\sim& (3,2,1,1/6), ~~~~~ d_L^c \sim (\overline 3, 1,1,1/3) \\
Q^c = (h^c,u^c)_L &\sim& (\overline 3, 1,2,-1/6), ~~~~~ h_L \sim (3,1,1,-1/3),
\end{eqnarray}
one bidoublet
\begin{equation}
\eta = \left( \begin{array} {c@{\quad}c} \eta_1^0 & \eta_2^+ \\ \eta_1^-
& \eta_2^0 \end{array} \right) \sim (1,2,2,0),
\end{equation}
six copies of
\begin{eqnarray}
\Phi_L = (\phi_L^0, \phi_L^-) &\sim& (1,2,1,-1/2), \\ \Phi_R = (\phi_R^+,
\phi_R^0) &\sim& (1,1,2,1/2),
\end{eqnarray}
three copies of
\begin{equation}
\Phi_L^c \sim (1,2,1,1/2), ~~~~~ \Phi_R^c \sim (1,1,2,-1/2),
\end{equation}
and six singlets
\begin{equation}
N \sim (1,1,1,0).
\end{equation}
Under SO(10), $Q$, $Q^c$, $\Phi_L$, and $\Phi_R$ belong to the {\bf 16},
$d^c$, $h$, and $\eta$ belong to the {\bf 10}, $\Phi_L^c$ and $\Phi_R^c$
belong to the {\bf 16*}.  Anomaly cancellation at the $\rm SU(3) \times
SU(2)_L \times SU(2)_R \times U(1)$ level is assured because of the
anomaly-free combinations $Q + Q^c + \Phi_L + \Phi_R$, $d^c + h$,
$\Phi_L + \Phi_L^c$, and $\Phi_R + \Phi_R^c$.  The left-right gauge
symmetry is broken spontaneously at $\rm M_R$ by the nonzero vacuum
expectation values of $\Phi_R$ and $\Phi_R^c$.  The scale of soft
supersymmetry breaking is assumed to coincide with $\rm M_R$.  The
effective particle content at the electroweak energy scale is assumed
to be that of the (nonsupersymmetric) standard $\rm SU(3) \times
SU(2)_L \times U(1)_Y$ model with two Higgs doublets.

Consider now the evolution of the gauge couplings to two-loop order.
Generically,
\begin{equation}
\mu {{\partial \alpha_i (\mu)} \over {\partial \mu}} = {1 \over {2\pi}}
\left( b_i + {b_{ij} \over {4\pi}} \alpha_j (\mu) \right) \alpha_i^2
(\mu),
\end{equation}
where $\alpha_i \equiv g_i^2/4\pi$ and $b_i$, $b_{ij}$ are constants
determined by the particle content contributing to $\alpha_i$.  The
initial conditions are set at $M_Z = 91.187 \pm 0.007$ GeV\cite{19} by
the experimental values $\alpha^{-1} = 127.9 \pm 0.1$,\cite{20}
$\sin^2 \theta_W = 0.2321 \pm 0.0006$,\cite{21} and $\alpha_S = 0.120
\pm 0.006 \pm 0.002$.\cite{22}  Hence
\begin{equation}
\alpha_S^{-1} (M_Z) = 8.33 \begin{array} {c} +0.60 \\ -0.52 \end{array} ~,
{}~~~ \alpha_L^{-1} (M_Z) = 29.69 \pm 0.10, ~~~ \alpha_Y^{-1} (M_Z) =
98.21 \pm 0.15.
\end{equation}
At $\rm M_R$, the matching conditions of the gauge couplings are
\begin{equation}
\alpha_L^{-1} = \alpha_R^{-1}, ~~~ \alpha_Y^{-1} = \alpha_R^{-1} +
\alpha_X^{-1},
\end{equation}
where $\alpha_X$ refers to the U(1) gauge coupling of the left-right
symmetry.  Above $\rm M_R$, $\alpha_L$ and $\alpha_R$ will evolve
together identically.

In the one-loop approximation, below $\rm M_R$,
\begin{eqnarray}
b_S &=& -11 + {4 \over 3} (3) ~=~ -7, \\
b_L &=& - {22 \over 3} + {4 \over 3} (3) + {1 \over 6} (2) ~=~ -3, \\
b_Y &=& {20 \over 9} (3) + {1 \over 6} (2) ~=~ 7,
\end{eqnarray}
whereas above $\rm M_R$,
\begin{eqnarray}
b_S &=& -9 + 2(3) + n_h = 0, \\
b_{LR} &=& -6 + 2(3) + n_{22} + n_\phi = 4, \\
{3 \over 2} b_X &=& 2(3) + 3n_\phi + n_h = 18,
\end{eqnarray}
where $n_{22} = 1$, $n_h = 3$, and $n_\phi = 3$ refer respectively to the one
bidoublet $\eta$, the three copies of $h + d^c$, and the three copies of
$\Phi_L + \Phi_L^c + \Phi_R + \Phi_R^c$, and the factor 3/2 for
$b_X$ comes from the normalization of the U(1)$_X$ coupling within SO(10).
Assuming that $\alpha_S^{-1} = \alpha_{LR}^{-1} = (3/2) \alpha_X^{-1}$ at
the unification scale $\rm M_U$ and neglecting the two-loop coefficients
$b_{ij}$, Eq. (8) can be solved for $\rm M_R$ and $\rm M_U$, {\it i.e.}
\begin{equation}
\ln {{\rm M_R} \over M_Z} = {\pi \over 4} \left[ 3 \alpha^{-1} (M_Z)
\{ 1 - 5 \sin^2 \theta_W (M_Z) \} + 7 \alpha_S^{-1} (M_Z) \right]
< 1.66,
\end{equation}
and
\begin{equation}
\ln {{\rm M_U} \over M_Z} = {\pi \over 2} \left[ \alpha^{-1} (M_Z)
\sin^2 \theta_W (M_Z) - \alpha_S^{-1} (M_Z) \right] > 32.45.
\end{equation}
Hence $\rm M_R < 480$ GeV and $\rm M_U > 1.1 \times 10^{16}$ GeV.  The
upper bound of $\rm M_U$ is $1.9 \times 10^{16}$ GeV, corresponding to
$\rm M_R = M_Z$.  Note that these results are identical to those of a
recently proposed extension of the
conventional supersymmetric left-right model\cite{18} because each
corresponding $b_i$ above $\rm M_R$ differs by the same amount, {\it i.e.}
one.

The allowed parameter space opens up more in two loops.  Using\cite{23}
\begin{equation}
b_{ij} = \left( \begin{array} {c@{\quad}c@{\quad}c} -26 & {9 \over 2} &
{11 \over 10} \\ 12 & 8 & {6 \over 5} \\ {44 \over 5} & {18 \over 5} &
{104 \over 25} \end{array} \right)
\end{equation}
for $\alpha_S^{-1}$, $\alpha_L^{-1}$, and (3/5)$\alpha_Y^{-1}$ below
$\rm M_R$, and
\begin{equation}
b_{ij} = \left( \begin{array} {c@{\quad}c@{\quad}c} 48 & 9 & 3 \\ 24 &
49 & {15 \over 2} \\ 24 & {45 \over 2} & {45 \over 2} \end{array} \right)
\end{equation}
for $\alpha_S^{-1}$, $\alpha_{LR}^{-1}$, and (3/2)$\alpha_X^{-1}$ above
$\rm M_R$, and solving Eq.~(8) numerically with the proper boundary
conditions at $\rm M_U$:
\begin{equation}
\alpha_U^{-1} - {2 \over {3\pi}} = \alpha_S^{-1} - {1 \over {4\pi}} =
\alpha_{LR}^{-1} - {1 \over {6\pi}} = {3 \over 2} \alpha_X^{-1},
\end{equation}
it is found that
\begin{equation}
\rm 1.0 \times 10^{16}~GeV < M_U < 2.3 \times 10^{16}~GeV
\end{equation}
with
\begin{equation}
{\rm 1.3~TeV > M_R} > M_Z.
\end{equation}
As an example, Fig. 1 shows the case with $\rm M_R = 1$ TeV and $\rm M_U =
1.1 \times 10^{16}$ GeV for the values $\alpha^{-1} (M_Z) = 127.9$,
$\sin^2 \theta_W (M_Z) = 0.2316$, and $\alpha_S (M_Z) = 0.112$.

In the original model,\cite{1} there are three copies of the {\bf 27}
representations of E$_6$, {\it i.e.} three copies of the {\bf 16},
{\bf 10}, and {\bf 1} representations of SO(10).  To arrive at the
present model, two bidoublets (belonging to the {\bf 10}) are removed,
but three copies of $\Phi_L + \Phi_R + \Phi_L^c + \Phi_R^c$ (belonging
to the {\bf 16} + {\bf 16*}) and three more singlets are added.  This
assumed modification at low energies is what makes this model
unifiable despite having $\rm M_R$ at about 1 TeV.  At the unification
scale $\rm M_U$, there are presumably at least six copies of {\bf 27}
and three copies of {\bf 27*}.  The missing components are assumed to
be superheavy with masses of order $\rm M_U$.

The interactions of this model at the $\rm SU(3) \times SU(2)_L \times
SU(2)_R \times U(1)$ level are assumed to obey a discrete $\rm Z_4 \times
Z_2$ symmetry, under which the various superfields transform as given
in Table 1.  Consider first only those terms in the superpotential
involving the quarks.  Only three are allowed:
\begin{eqnarray}
Q Q^c \eta &=& d h^c \eta_1^0 - u h^c \eta_1^- + u u^c \eta_2^0 - d u^c
\eta_2^+, \\ Q d^c \Phi_{L(1)} &=& d d^c \phi_{L(1)}^0 - u d^c
\phi_{L(1)}^-, \\ h Q^c \Phi_{R(1)} &=& h h^c \phi_{R(1)}^0 -
h u^c \phi_{R(1)}^+.
\end{eqnarray}
This means that the exotic heavy quark $h$ gets its mass from the vacuum
expectation value $\langle \phi_{R(1)}^0 \rangle$, the ordinary quarks $u$ and
$d$ get their masses from $\langle \eta_2^0 \rangle$ and $\langle \phi_{L(1)}^0
\rangle$ respectively, and the fermionic components of the $\rm SU(2)_L$
doublet $(\eta_1^0, \eta_1^-)$ and singlet $\phi^+_{R(1)}$ can be
identified as ordinary leptons,\cite{1} such as $(\nu_\tau, \tau^-)$ and
$\tau^+$.  Consequently, $h$ picks up a nonzero lepton number $L = 1$ and
since $W_R^-$ converts $h^c$ to $u^c$, it also has $L = 1$.  This prevents
$d-h$ as well as $W_L - W_R$ mixing, and since only one scalar vacuum
expectation value contributes to the mass of each quark species, the
absence of tree-level flavor-changing neutral-current interactions is
also assured.

Consider next the terms involving $\Phi_L^c$ and $\Phi_R^c$.  There are
three quadratic terms:
\begin{equation}
\Phi_{L(4,5,6)} \Phi^c_{L(1,2,3)}, ~~~~
\Phi_{R(4)} \Phi^c_{R(1)}, ~~~~
\Phi_{R(5,6)} \Phi^c_{R(2,3)}.
\end{equation}
This means that $\Phi_{L(4,5,6)} + \Phi^c_{L(1,2,3)} + \Phi_{R(4,5,6)} +
\Phi^c_{R(1,2,3)}$ can be assumed heavy with masses of order $\rm M_R$.
There are also three cubic terms:
\begin{equation}
\Phi_{L(2,3)} \Phi^c_{L(1,2,3)} N_L, ~~~~
\Phi_{R(1)} \Phi^c_{R(1)} N_R, ~~~~
\Phi_{R(2,3)} \Phi^c_{R(2,3)} N_R.
\end{equation}
This means that $N_R$ should have $L = 0$ whereas $\Phi_{L(2,3)}$ can be
assigned $L = 1$, $\phi^+_{R(2,3)}$ and $N_L$ assigned $L = -1$.  However, the
quadratic terms $N_L N_L$ and $N_R N_R$ are also allowed.  Hence additive
lepton number is explicitly violated by the $N_L$ Majorana mass terms, but they
are the sole source of this violation, as is often the case when the
standard model is extended to include neutral singlet leptons.

The remaining allowed terms in the superpotential all involve the bidoublet:
\begin{eqnarray}
\eta \eta N_L &=& \eta_1^0 \eta_2^0 N_L - \eta_1^- \eta_2^+ N_L, \\
\Phi_{L(1)} \Phi_{R(1)} \eta &=& \phi^-_{L(1)} \phi^+_{R(1)} \eta_1^0 -
\phi^0_{L(1)} \phi^+_{R(1)} \eta_1^- + \phi^0_{L(1)} \phi^0_{R(1)} \eta_2^0
- \phi^-_{L(1)} \phi^0_{R(1)} \eta_2^+,
\end{eqnarray}
and
\begin{equation}
\Phi_{L(4,5,6)} \Phi_{R(2,3)} \eta.
\end{equation}
This means that the fermionic component of $\eta_1^0$ ({\it i.e.} $\nu_\tau$)
gets a seesaw mass through its coupling to $N_L$ via $\langle \eta_2^0
\rangle$ and the $N_L$ Majorana mass.  Since the $\tau$ lepton is
identified as the fermionic components of $\eta_1^-$ and $\phi^+_{R(1)}$,
it gets a mass via $\langle \phi^0_{L(1)} \rangle$.
The particle content of this model under $\rm SU(3) \times SU(2)_L \times
U(1)_Y$ is given in Table 2 together with the baryon number $B$, lepton
number $L$, and $R$ parity of the fermions $R_f = (-1)^{1+3B+L}$.  Strictly
speaking, because $N_L$ has a Majorana mass, lepton number is conserved only
multiplicatively.

The electron and muon are identified in this model as the fermionic
components of $\phi^-_{L(2,3)}$ and $\phi^+_{R(2,3)}$, whereas $\nu_e$
and $\nu_\mu$ are equated with the fermionic components of $\phi^0_{L(2,3)}$.
The latter are coupled to $N_L$ via $\langle (\phi^c_L)^0 \rangle$, hence
they acquire seesaw masses in the same way as $\nu_\tau$ and all three
neutrinos may mix with one another.  On the other hand, $e$ and $\mu$
are massless at tree level.  To see how they acquire radiative masses,
note that the soft supersymmetry-breaking term
\begin{equation}
\Phi_{L(2,3)} \Phi_{R(2,3)} \tilde \eta = \phi^-_{L(2,3)} \phi^+_{R(2,3)}
\overline {\eta_2^0} - \phi^0_{L(2,3)} \phi^+_{R(2,3)} \eta_2^- +
\phi^0_{L(2,3)} \phi^0_{R(2,3)} \overline {\eta_1^0} - \phi^-_{L(2,3)}
\phi^0_{R(2,3)} \eta_1^+
\end{equation}
involving only the scalar fields is allowed by the discrete $\rm Z_4 \times
Z_2$ symmetry.  Since $\phi^-_{L(2,3)}$ and $\phi^+_{R(2,3)}$ are now
identified as the scalar supersymmetric partners of $e$ and $\mu$, the
radiative mechanism\cite{24,25} of gaugino exchange allows $e$ and $\mu$
to become massive via $\langle \eta_2^0 \rangle$.  Note that there is
mixing between $\phi^-_{L(2,3)}$ and $\eta_1^+$ via $\langle \phi^0_{R(2,3)}
\rangle$, hence $e$ and $\mu$ also mix with $\tau$ through radiative
corrections.  In addition, because of the $\Phi_{L(4,5,6)} \Phi_{R(2,3)}
\eta$ term, flavor-changing leptonic processes are possible at the TeV
energy scale.  Phenomenological details will be given elsewhere.

In summary, the proposed supersymmetric left-right model of this paper has
the following interesting features.  (1) Its particle content is chosen from
six copies of the {\bf 27} and three copies of the {\bf 27*} of E$_6$,
consistent with the possibility that it is of superstring origin.  (2)
It is unifiable at around $10^{16}$ GeV even though the $\rm SU(3) \times
SU(2)_L \times SU(2)_R \times U(1)$ gauge symmetry is realized at the TeV
energy scale with $g_L = g_R$.  (3) The problem of requiring two or more
scalar bidoublets for realistic quark masses in the conventional left-right
model is circumvented because the $\rm SU(2)_R$ doublet is not $(u,d)_R$,
but $(u,h)_R$ where $h$ is heavy and has lepton number $L = 1$.  This allows
the model to be free of tree-level flavor-changing neutral currents at the
electroweak energy scale.  (4) The $\tau$ lepton gets a mass spontaneously
as usual, but $e$ and $\mu$ acquire radiative masses through
gaugino exchange.  (5) All three neutrinos have small seesaw masses through
their couplings to neutral gauge singlets with large Majorana masses.
Additive lepton number $L$ is explicitly violated but multiplicative lepton
number $(-1)^L$ is preserved.  (6) The structure of this model is
naturally maintained with a discrete $\rm Z_4 \times Z_2$ symmetry
which is spontaneously broken down to a generalization of $R$ parity.
(7) At the TeV energy scale, there will be many unique manifestations of
this model.  For example, the $W_R$ vector gauge boson here has a nonzero
lepton number and negative $R$ parity, hence its final decay product must
contain at least a lepton as well as the LSP (lightest supersymmetric
particle).
\vspace{0.3in}
\begin{center} {ACKNOWLEDGEMENT}
\end{center}

This work was supported in part by the U. S. Department of Energy under
Grant No. DE-FG03-94ER40837.

\newpage
\bibliographystyle{unsrt}

\vspace{0.3in}

\begin{center} {FIGURE CAPTION}
\end{center}

\noindent Fig.~1.  Evolution of $\alpha_i^{-1}$ with $\rm M_R = 1~TeV$ and
$\rm M_U = 1.1 \times 10^{16}~GeV$.

\newpage
\begin{table}
\begin{center}
\begin{math}
\begin{array} {|c|c|c|} \hline
\rm Superfield & \rm Z_4 & \rm Z_2 \\
\hline
Q & 1 & + \\
Q^c & -i & + \\
d^c & 1 & + \\
h & -1 & + \\
\hline
\eta & i & + \\
\hline
\Phi_{L(1)} & 1 & + \\
\Phi_{L(2,3)} & i & - \\
\Phi_{L(4,5,6)} & -i & - \\
\hline
\Phi_{R(1)} & -i & + \\
\Phi_{R(2,3)} & 1 & - \\
\Phi_{R(4)} & i & - \\
\Phi_{R(5,6)} & -1 & + \\
\hline
\Phi^c_{L(1,2,3)} & i & - \\
\hline
\Phi^c_{R(1)} & -i & - \\
\Phi^c_{R(2,3)} & -1 & + \\
\hline
N_{L(1,2,3)} & -1 & + \\
N_{R(1,2,3)} & -1 & - \\
\hline
\end{array}
\end{math}
\end{center}
\caption {Transformation properties of the various superfields of this
model under $\rm Z_4 \times Z_2$.}
\end{table}

\newpage
\begin{table}
\begin{center}
\begin{math}
\begin{array} {|c|c|c|c|c|} \hline
\rm Superfield & \rm SU(3) \times SU(2) \times U(1) & B & L & R_f \\
\hline
(u,d) & (3,2,1/6) & 1/3 & 0 & + \\
d^c & (\overline 3, 1, 1/3) & -1/3 & 0 & + \\
u^c & (\overline 3, 1, -2/3) & -1/3 & 0 & + \\
\hline
h & (3, 1, -1/3) & 1/3 & 1 & - \\
h^c & (\overline 3, 1, 1/3) & -1/3 & -1 & - \\
\hline
(\eta_1^0, \eta_1^-) & (1,2,-1/2) & 0 & 1 & + \\
(\phi^0_{L(2,3)}, \phi^-_{L(2,3)}) & (1,2,-1/2) & 0 & 1 & + \\
\phi^+_R & (1,1,1) & 0 & -1 & + \\
(\phi_R^c)^- & (1,1,-1) & 0 & 1 & + \\
N_L & (1,1,0) & 0 & -1 & + \\
\hline
\phi_R^0, (\phi_R^c)^0, N_R & (1,1,0) & 0 & 0 & - \\
\Phi_{L(1,4,5,6)} & (1,2,-1/2) & 0 & 0 & - \\
(\eta_2^+, \eta_2^0), \Phi_L^c & (1,2,1/2) & 0 & 0 & - \\
\hline
\end{array}
\end{math}
\end{center}
\caption {Transformation properties of the various superfields of this
model under $\rm SU(3) \times SU(2) \times U(1)$, $B$, $L$, and $R_f$.}
\end{table}

\end{document}